

\input harvmac

\overfullrule=0pt


\def\J{{\scriptscriptstyle J}}

\def\Q{{\scriptscriptstyle Q}}


\def\CA{{\cal A}}

\def\CM{{\cal M}}

\def\CO{{\cal O}}


\def\d{\delta}

\def\o{\sigma}

\def\th{\theta}
\def\u{\mu}
\def\v{\nu}


\def\aS{\alpha_s}
\def\bar#1{\overline{#1}}
\def\bbar{{\overline b}}
\def\bbbaroctet{b\bbar[^3S_1^{(8)}]}
\def\Bbar{{\overline B}}
\def\Br{{\rm Br}}
\def\cbar{{\overline c}}
\def\ccbaroctet{c\cbar[^3S_1^{(8)}]}
\def\ccdot{\hbox{\kern-.1em$\cdot$\kern-.1em}}
\def\chibJ{\chi_{b\J}}
\def\chicJ{\chi_{c\J}}
\def\chiQJ{\chi_{\Q\J}}

\def\Dbar{{\overline D}}

\def\GeV{\>\, \rm GeV}
\def\gfive{\gamma^5}
\def\gS{g_s}
\def\gtap{\raise.3ex\hbox{$>$\kern-.75em\lower1ex\hbox{$\sim$}}}
\def\Jpsi{J/\psi}

\def\longmatrixelement{\CM_8(\psi_Q)}
\def\ltap{\raise.3ex\hbox{$<$\kern-.75em\lower1ex\hbox{$\sim$}}}

\def\Mb{M_b}

\def\Mc{M_c}

\def\MQ{M_\Q}

\def\nf{n_f}
\def\octet{^3\hskip-0.2 em S_1^{(8)}}

\def\pbar{{\overline{p}}}

\def\psiQ{\psi_\Q}

\def\Pslash{P\hskip-0.5em /}
\def\pT{p_\perp}
\def\qbar{{\overline q}}
\def\Qbar{{\overline Q}}
\def\QCD{{\scriptscriptstyle QCD}}
\def\QQbaroctet{Q\Qbar[^3S_1^{(8)}]}
\def\QQbarocteth{Q\Qbar^{(h)}[^3S_1^{(8)}]}
\def\qslash{q\hskip-0.5em /}
\def\shat{\hat{s}}
\def\slash#1{#1\hskip-0.5em /}
\def\sp{\>\>}

\def\that{\hat{t}}
\def\ubar{\bar{u}}
\def\uhat{\hat{u}}
\def\vbar{\bar{v}}
\def\vb{v_b}
\def\vc{v_c}
\def\vQ{v_\Q}


\def\half{{1 \over 2}}


\newdimen\pmboffset
\pmboffset 0.022em
\def\oldpmb#1{\setbox0=\hbox{#1}%
 \copy0\kern-\wd0 \kern\pmboffset\raise
 1.732\pmboffset\copy0\kern-\wd0 \kern\pmboffset\box0}


%
%
\def\appendix#1#2{\global\meqno=1\global\subsecno=0\xdef\secsym{\hbox{#1.}}
\bigbreak\bigskip\noindent{\bf Appendix. #2}\message{(#1. #2)}
\writetoca{Appendix {#1.} {#2}}\par\nobreak\medskip\nobreak}


\nref\BraatenYuanI{E. Braaten and T.C. Yuan, Phys. Rev. Lett. {\bf 71} (1993)
 1673.}
\nref\BCY{E. Braaten, K. Cheung and T.C. Yuan, Phys. Rev. {\bf D48} (1993)
 4230.}
\nref\Chen{Y.-Q. Chen, Phys. Rev. {\bf D48} (1993) 5181.}
\nref\BraatenYuanII{E. Braaten and T.C. Yuan, Phys. Rev. {\bf D50} (1994)
 3176.}
\nref\Yuan{T.C. Yuan, Phys. Rev. {\bf D50} (1994) 5664.}
\nref\BDFM{E. Braaten, M.A. Doncheski, S. Fleming and M.L. Mangano,
 Phys. Lett. {\bf B333} (1994) 548.}
\nref\Roy{D.P. Roy and K. Sridhar, Phys. Lett. {\bf B339} (1994) 141.}
\nref\Cacciari{M. Cacciari and M. Greco, Phys. Rev. Lett. {\bf 73} (1994)
 1586.}
\nref\Psidata{The CDF collaboration, Fermilab-Conf-94/136-E (1994),
 unpublished.}
\nref\Cho{P. Cho, S. Trivedi and M. Wise, Phys. Rev. {\bf D51} (1995) 2039.}
\nref\Close{F.E. Close, Phys. Lett. {\bf B342} (1995) 369.}
\nref\ChoWise{P.Cho and M. Wise, Phys. Lett. {\bf B346} (1995) 129.}
\nref\BraatenFleming{E. Braaten and S. Fleming, NUHEP-TH-94-26 (1994),
  unpublished.}
\nref\Upsilondata{The CDF collaboration, Fermilab-Conf-94/221-E (1994),
 unpublished.}
\nref\Bodwin{G.T. Bodwin, E. Braaten and G.P. Lepage, Phys. Rev. {\bf D51}
(1995) 1125.}
\nref\Caswell{W.E. Caswell and G.P. Lepage, Phys. Lett. {\bf B167} (1986)
 437.}
\nref\Lepage{G.P. Lepage, L. Magnea, C. Nakhleh, U. Magnea and K.
 Hornbostel, Phys. Rev. {\bf D46} (1992) 4052.}
\nref\Quarkonia{For reviews of quarkonia potential model calculations, see
 {\it Quarkonia}, edited by W. Buchm\"uller, (North-Holland, Amsterdam)
 1992.}
\nref\Baier{R. Baier and R. R\"uckl, Z. Phys. C {\bf 19} (1983) 251.}
\nref\Gastmans{R. Gastmans, W. Troost and T.T. Wu, Nucl. Phys. {\bf B291}
 (1987) 731.}
\nref\Humpert{B. Humpert, Phys. Lett. {\bf B184} (1987) 105.}
\nref\Schuler{G.A. Schuler, CERN-TH.7170/94 (1994), unpublished, and
 references therein.}
\nref\Kuhn{J. H. K\"uhn, J. Kaplan and E. G. O. Safiani, Nucl. Phys. {\bf
 B157} (1979) 125.}
\nref\Guberina{B. Guberina, J.H. K\"uhn, R.D. Peccei and R. R\"uckl, Nucl.
 Phys. {\bf B174} (1980) 317.}
\nref\Yang{T.D. Lee, {\it Particle Physics and Introduction to Field Theory},
 (Harwood Academic Publishers, Chur, Switzerland) 1988.}
\nref\Mertig{R. Mertig, M. B\"ohm and A. Denner, Comp. Phys. Comm. {\bf 64}
 (1991) 345.}
\nref\Quigg{E.J. Eichten and C. Quigg, Fermilab-Pub-95/045-T (1995),
 unpublished.}
\nref\Buchmuller{W. Buchm\"uller and S.-H. H. Tye, Phys. Rev. {\bf D24} (132)
 1981.}
\nref\CLEO{The CLEO collaboration, CLEO CONF 94-6 (1994), unpublished.}
\nref\Bauer{G. Bauer representing the CDF collaboration, Talk presented at
the $p\pbar$ 95 workshop, May 1995.}
\nref\forthcoming{P. Cho, in preparation.}


\nfig\chiQJproduction{Typical Feynman graphs which contribute to $\chiQJ$
production at $O(\aS^3)$ through the (a) color-singlet and (b) color-octet
mechanisms.  The short-distance scattering collisions respectively create
$Q\Qbar$ pairs in $^3P_J^{(1)}$ and $\octet$ configurations.  The $Q\Qbar$
pairs hadronize at long distances into $\chiQJ$ bound states.}
\nfig\qqbaroctet{Color-octet diagram which mediates $q\qbar \to \psiQ$
production at $O(\aS^2)$.}
\nfig\goctet{Lowest order contribution to the $g^* \to \QQbaroctet$
subamplitude in eqn.~2.5.}
\nfig\ggoctet{Color-octet diagrams which mediate $gg \to
\psiQ$ production at $O(\aS^2)$.  The sum of these graphs vanishes when both
incoming gluons are on-shell.}
\nfig\octetgraphs{Color-octet diagrams which mediate (a) $q\qbar \to
\psiQ g$, (b) $g q \to \psiQ q$ and (c) $gg \to \psiQ g$ scattering at
$O(\aS^3)$.  The shaded circles appearing in these graphs represent the
$q\qbar \to \psiQ$ and $gg \to \psiQ$ amplitudes pictured in \qqbaroctet\ and
\ggoctet.}
\nfig\UpsilononeSxsect{Theoretical transverse momentum differential cross
section for $\Upsilon(1S)$ production at the Tevatron in the rapidity interval
$|y| \le 0.4$ compared against preliminary CDF data.  The dashed curve
depicts the color-singlet contribution which includes direct $\Upsilon(1S)$
production as well as radiative feeddown from $\chibJ(1P)$ and $\chibJ(2P)$
states.  The dot-dashed curve illustrates the color-octet contribution to
$\Upsilon(1S)$ production.  The solid curve equals the sum of the color-singlet
and color-octet contributions and represents the total theoretical differential
cross section.  The $\Upsilon(1S)$ cross section predicted by the high energy
gluon fragmentation approximation is shown by the dotted curve.}
\nfig\UpsilontwoSxsect{Theoretical transverse momentum differential cross
section for $\Upsilon(2S)$ production at the Tevatron in the rapidity interval
$|y| \le 0.4$ compared against preliminary CDF data.  The curves in this
figure are labeled the same as those in \UpsilononeSxsect.  The dashed
color-singlet cross section includes $\Upsilon(2S)$ production and radiative
feeddown from $\chibJ(2P)$.}
\nfig\UpsilonthreeSxsect{Theoretical transverse momentum differential cross
section for $\Upsilon(3S)$ production at the Tevatron in the rapidity interval
$|y| \le 0.4$ compared against preliminary CDF data.  The curves in this figure
are labeled the same as those in \UpsilononeSxsect.  The dashed color-singlet
cross section includes $\Upsilon(3S)$ production and radiative feeddown from
$\chibJ(3P)$.  The measured $n=2$ radiative branching ratios
$\Br(\chibJ(2P) \to \Upsilon(2S) + \gamma)$ have been used as estimates for
their unknown $n=3$ counterparts $\Br(\chibJ(3P) \to \Upsilon(3S) + \gamma)$.}
\nfig\Jpsixsect{Theoretical transverse momentum differential cross
section for $\Jpsi$ production at the Tevatron in the pseudorapidity
interval $|\eta| \le 0.6$ compared against preliminary CDF data.  The
dashed curve depicts the color-singlet contribution which includes direct
$\Jpsi$ production and radiative feeddown from $\chicJ$ states.  The
dot-dashed curve illustrates the color-octet contribution to $\Jpsi$
production.  The solid curve equals the sum of the color-singlet and
color-octet contributions and represents the total theoretical differential
cross section.  The $\Jpsi$ cross section predicted by the high energy gluon
fragmentation approximation is shown by the dotted curve.}
\nfig\Psipxsect{Theoretical transverse momentum differential cross section for
$\psi'$ production at the Tevatron in the pseudorapidity interval
$|\eta| \le 0.6$ compared against preliminary CDF data.  The curves in this
figure are labeled the same as those in \Jpsixsect. The dashed color-singlet
cross section includes only direct $\psi'$ production.}
\nfig\Transratio{Ratio $R_{\rm trans}$ of transversely aligned and total
$\QQbaroctet$ differential cross sections plotted as a function of $\pT$.
The solid and dot-dashed curves illustrate $R_{\rm trans}$ for $b\bbar[\octet]$
and $c\cbar[\octet]$ production respectively.}
%


\def\CITTitle#1#2#3{\nopagenumbers\abstractfont
\hsize=\hstitle\rightline{#1}
\vskip 0.4in\centerline{\titlefont #2} \centerline{\titlefont #3}
\abstractfont\vskip .4in\pageno=0}

\CITTitle{{\baselineskip=12pt plus 1pt minus 1pt
  \vbox{\hbox{CALT-68-1988}\hbox{DOE RESEARCH AND}\hbox{DEVELOPMENT
  REPORT}}}}
{Color-Octet Quarkonia Production}{}
\centerline{
  Peter Cho\footnote{$^1$}{Work supported in part by a DuBridge Fellowship and
  by the U.S. Dept. of Energy under DOE Grant no. DE-FG03-92-ER40701.}
  and Adam K. Leibovich\footnote{$^2$}{Work supported in part by
  the U.S. Dept. of Energy under DOE Grant no. DE-FG03-92-ER40701.}}
\centerline{Lauritsen Laboratory}
\centerline{California Institute of Technology}
\centerline{Pasadena, CA  91125}

\vskip .2in
\centerline{\bf Abstract}
\bigskip

	Gluon fragmentation represents the dominant source of high energy
prompt quarkonia at hadron colliders.  Fragmentation approximations break down,
however, when a \break quarkonium's transverse momentum becomes comparable to
its mass.  In this paper, we identify a large class of color-octet diagrams
that mediate quarkonia production at all energies and reduce to the dominant
set of gluon fragmentation graphs in the high $\pT$ limit.  They contribute to
quarkonia differential cross sections at the same order as color-singlet
diagrams and bring theoretical predictions for Upsilon and Psi production at
the Tevatron into agreement with experimental measurements.  Using recent CDF
data, we extract numerical values for bottomonia and charmonia color-octet
matrix elements which are consistent with NRQCD scaling rules.  We also find
that quarkonia generated via the color-octet mechanism are strongly polarized
at low as well as high energies.  Spin alignment measurements can thus test
the color-octet quarkonia production picture.

\Date{5/95}

\newsec{Introduction}

	The theory of quarkonia production has recently undergone a
number of significant developments.  Progress in this area was stimulated by
the discovery of Braaten and Yuan in 1993 that parton fragmentation represents
the dominant source of prompt quarkonia at high energies \BraatenYuanI.  These
authors realized that the fragmentation functions which specify the
probability for gluons and heavy quarks to hadronize into quarkonia bound
states can be calculated starting from first principles in QCD.  Over the past
few years, the most important fragmentation functions for S and P-wave
quarkonia have been computed to lowest order \refs{\BraatenYuanI{--}\Yuan}.
They have subsequently been utilized to investigate a range of issues related
to quarkonium phenomenology.

	One of the most interesting applications of this fragmentation
work has been to the study of prompt charmonium production at the Tevatron
\refs{\BDFM{--}\Cacciari}.  Comparison between theoretical predictions and
experimental measurements of $\Jpsi$ and $\psi'$ differential cross sections
clearly demonstrate that fragmentation dominates over all other charmonium
creation mechanisms at large transverse momenta.  Orders of magnitude
discrepancies between cross section computations and observations are
significantly reduced when gluon and charm fragmentation processes are taken
into account.  The predicted prompt $\Jpsi$ rate then qualitatively agrees
with recent CDF data \Psidata.  In the case of $\psi'$ production, the
fragmentation results substantially improve upon earlier differential cross
section predictions.  But they still underestimate the number of $\psi'$'s
observed at the Tevatron by more than a factor of 30.   This large discrepancy
between theory and experiment indicates that some important production
mechanism beyond the simplest $g \to \psi'$ and $c \to \psi'$ fragmentation
processes needs to be included.  A number of possible resolutions to the
$\psi'$ problem have been suggested \refs{\Cho{--}\BraatenFleming}, and
experimental studies which may confirm or reject these proposals are underway.

	Although the fragmentation picture of quarkonium production has
provided valuable insight, its validity domain is restricted to high energies.
The approximations that enter into fragmentation function computations break
down when a quarkonium's energy becomes comparable to its mass.  Fragmentation
predictions for charmonium differential cross sections are therefore
unreliable at low transverse momenta.  Similarly, fragmentation results for
bottomonium production are untrustworthy throughout the $\pT < 15 \GeV$ region
where Tevatron data exists.  Yet even in this low energy regime, Upsilon cross
section measurements significantly disagree with existing theoretical
predictions \Upsilondata.

	The discrepancies in the Upsilon sector provide the primary motivation
for the work which we present in this paper.  Building upon several recently
developed ideas in heavy quarkonium physics, we identify a large class of
color-octet diagrams that mediate quarkonia production at all energies and
reduce to the dominant set of gluon fragmentation graphs in the high $\pT$
limit.  These diagrams enter at the same order into direct production processes
as previously considered color-singlet graphs and must be included for
consistency.  As we shall see, the addition of color-octet contributions to
Upsilon production computations significantly improves agreement between
theory and experiment.

	Our article is organized as follows.  In section 2, we discuss
the short and long distance factorization of quarkonia production amplitudes.
We then calculate the leading color-octet contributions to quarkonia
differential cross sections.  In section 3, we apply the color-octet cross
sections to the study of Upsilon and Psi production at the Tevatron.
Using CDF data, we extract numerical values for bottomonia and
charmonia color-octet matrix elements which we compare with NRQCD scaling
relations in section 4.  We next analyze the spin alignment of quarkonia
generated via the color-octet mechanism as a function of transverse momentum
in section 5.  Finally, we summarize our findings in section 6 and close with
some thoughts on directions for future work.

\newsec{Color octet contributions to quarkonia cross sections}

	The physics of quarkonia involves several different energy scales
that are separated by the small velocity $v$ of the two heavy constituents
inside $Q\Qbar$ bound states \Bodwin.  The first is set by the quark mass
$\MQ$ which fixes the distance range for $Q\Qbar$ creation and annihilation
processes.  The second is given by the heavy constituents' typical momentum
$\MQ v$ which is inversely proportional to the bound state's spatial size.
The kinetic energy $\MQ v^2$ represents a third important quarkonium energy
scale and determines the constituents' interaction time.  As $\MQ \to \infty$,
the heavy constituents' velocity $v \sim 1/\log \MQ$ tends towards zero, and
these three scales become widely separated:
\eqn\scales{(\MQ v^2)^2 \ll (\MQ v)^2 \ll \MQ^2.}
This hierarchy is well realized in the real world for bottomonia with
$v_b^2 \simeq 0.08$ and to a lesser extent for charmonia with
$v_c^2 \simeq 0.23.$

	The presence of these different distance scales renders the study
of quarkonia both interesting and challenging.  An effective field theory
formalism called Nonrelativistic Quantum Chromodynamics (NRQCD) has been
established to systematically keep track of this scale hierarchy
\refs{\Bodwin,\Caswell}.  It is based upon a double power series expansion in
the strong interaction fine structure constant $\aS=\gS^2/4\pi$ and the
small velocity parameter $v$.  Short distance processes which occur at energies
comparable to or greater than the heavy quark mass are perturbatively
computable within the effective theory.  Long distance effects are consolidated
into nonperturbative matrix elements whose values must be determined from
experiment or the lattice.   High and low energy interactions are thus
cleanly separated in NRQCD, and simple power counting rules methodically
determine the dominant contributions to various quarkonium processes \Lepage.
We will work within the NRQCD framework throughout this paper.

	Before the development of NRQCD, quarkonia were generally treated as
nonrelativistic bound states of a heavy quark and antiquark in a static gluon
field that sets up an instantaneous confining potential \Quarkonia.
Although this picture has enjoyed a remarkable degree of phenomenological
success, it fails to take into account gluons inside a quarkonium with
wavelengths much greater than the bound state's characteristic size.  The
presence of such low energy gluons implies that the heavy quark and antiquark
cannot always be regarded as residing in a color-singlet configuration.

	This shortcoming of the potential model approach is rectified in
NRQCD.   Dynamical gluons enter into Fock state decompositions of physical
quarkonium states.  For example, wavefunctions for S-wave orthoquarkonia
schematically appear in Coulomb gauge as
\eqn\psiwavefunc{\eqalign{
|\, \psiQ \rangle &= O(1) |\, Q\Qbar\,[{}^3S_1^{(1)}] \,\rangle
+ O(v) |\, Q\Qbar\, [ {}^3P_J^{(8)}] \,  g \rangle \cr
& \quad + O(v^2) |\, Q\Qbar\, [ {}^3S_1^{(1,8)}] \,  g g \rangle
+ O(v^2) |\, Q\Qbar\, [ {}^1S_0^{(8)}] \,  g \rangle
+ O(v^2) |\, Q\Qbar\, [ {}^3D_J^{(1,8)}] \,  g g \rangle + \cdots.}}
The angular momentum quantum numbers of the $Q\Qbar$ pairs within the various
Fock components are indicated in spectroscopic notation inside the square
brackets, and their color configurations are labeled by singlet
or octet superscripts.  S-wave quarkonia reactions proceed at lowest
order in the velocity expansion through the first Fock state in \psiwavefunc.
The $Q\Qbar$ pair in this $O(1)$ Fock component has precisely the same
angular momentum and color quantum numbers as the full quarkonium state.
The NRQCD description of $\psiQ$ production or annihilation
thus reduces to the familiar color-singlet model in the $v \to 0$ limit.

	P-wave orthoquarkonia processes are more complicated.  They are
mediated at leading order in the velocity expansion by the first two Fock
states in the $\chiQJ$ wavefunction
\eqn\chiwavefunc{|\, \chiQJ\rangle = O(1) |\, Q\Qbar\, [ {}^3P_J^{(1)}] \,
 \rangle + O(v) |\, Q\Qbar\, [ {}^3S_1^{(8)}] \,  g \rangle + \cdots.}
Contributions to $\chiQJ$ formation which involve the first color-singlet Fock
component are proportional to the squared derivative of the P-wave bound
state's wavefunction at the origin.  This quantity counts as $O(v^5)$ in the
NRQCD velocity expansion.  On the other hand, short distance creation of the
S-wave color-octet $Q\Qbar$ pair in the second Fock state of \chiwavefunc\
takes place at $O(v^3)$.  The subsequent long distance evolution of the
$\QQbaroctet$ pair into a colorless $\chiQJ$ hadron via the emission or
absorption of a soft gluon costs an additional power of $v$ in the amplitude
and $v^2$ in the rate.  So $\chiQJ$ production proceeds at $O(v^5)$ in both
the color-singlet and color-octet mechanisms \BraatenYuanII.

	It is useful to draw a picture which clarifies the difference
between these two quarkonium production mechanisms.  In
\chiQJproduction a, we illustrate a typical color-singlet Feynman graph that
creates a $\chiQJ$.  The $O(\aS^3)$ hard scattering forms a colorless
$Q\Qbar[ {}^3P_J^{(1)}]$ pair at a short distance scale.  The heavy quark and
antiquark fly out from the initial collision point in nearly parallel
directions and almost on mass shell.  After exchanging many soft gluons, the
color-singlet $Q\Qbar$ pair eventually hadronizes at a long distance into a
$\chiQJ$ bound state.   In \chiQJproduction b, the $O(\aS^3)$ high energy
collision creates a $Q\Qbar$ pair in a color-octet configuration.  Far away
{}from the collision point, the $\QQbaroctet$ pair emits a long wavelength
gluon which bleeds off its color but carries away virtually no energy or
momentum.  The heavy pair thus transforms into a colorless $\chiQJ$ quarkonium
state.

	Color-singlet quarkonium production has been studied for years.
Leading order color-singlet differential cross sections were calculated a
decade ago \refs{\Baier{--}\Humpert}, and some total cross section formulae
have been evaluated at next-to-leading order as well \Schuler.  In contrast,
color-octet contributions to quarkonium processes have only recently begun to
be considered.  While the latter mechanism may be less familiar than the
former, it is certainly not less important.  As can be seen in
\chiQJproduction, color-singlet and color-octet graphs arise at the same
order in $\aS$.  Moreover, the color-octet diagrams dominate at high
energies.  So a complete analysis of quarkonium production at hadron colliders
must include both mechanisms.

	Color-octet creation of S-wave orthoquarkonia starts at $O(\aS^2)$
via the $q\qbar \to \psiQ$ Feynman diagram shown in \qqbaroctet.  As indicated
by the dashed line in the figure, the total color-octet amplitude factorizes
into short and long distance pieces:
\eqn\qqbarpsiQamp{\CA\bigl(q\qbar \to \psiQ(nS)\bigr)_{\rm octet}
= \CA(q\qbar \to \QQbaroctet)_{\rm short \> distance} \times
\CA\bigl(\QQbaroctet \to \psiQ(nS)\bigr)_{\rm long \> distance}.}
The short distance part of the quark-antiquark fusion process lies to the left
of the dashed line in \qqbaroctet\ and can be calculated within perturbative
QCD.  On the other hand, the long distance component to the right of the
dashed line involves nonperturbative physics.  We will consider these two
factors in turn.

	To begin, we examine the subamplitude for virtual gluon formation of
a color-octet $Q\Qbar$ pair
\eqn\goctetampI{\CA \bigl(g_a^* \to \QQbaroctet_b  \bigr)_\mu =
\sum_{s_1,s_2} \sum_{i,j} \ubar\bigl( {P \over 2}+q;s_1\bigr)
\gS \gamma_\mu (T_a)^i_j v\bigl({P \over 2} - q;s_2 \bigr)
\langle \half s_1 ; \half s_2 | 1 S_z \rangle \>
\langle 3 i ; \bar{3} j | 8 b \rangle }
which is illustrated in \goctet.  The heavy quark and antiquark propagate
nearly on-shell with a large combined momentum $P$ and a small relative
momentum $q$.  The spin and color quantum numbers for the $Q\Qbar$ pair are
projected out by the sums over the $SU(2)$ and $SU(3)$ Clebsch-Gordan
coefficients $\langle \half s_1 ; \half s_2 | 1 S_z \rangle$ and
$\langle 3 i ; \bar{3} j | 8 b \rangle = \sqrt{2} (T_b)^j_i$.  The
subamplitude may be rewritten in the trace form
\eqn\goctetampII{\CA(g_a^* \to \QQbaroctet_b)_\mu = \sqrt{2} \gS
\Tr(T_a T_b) \Tr[ \gamma_\mu P_{1 S_z} (P;q) ] }
where
\eqn\spinprojopI{P_{S S_z}(P;q) = \sum_{s_1,s_2}
v\bigl({P \over 2} - q;s_2\bigr)
\ubar\bigl( {P \over 2}+q;s_1\bigr)
\langle \half s_1 ; \half s_2 | S S_z \rangle}
denotes a spin projection operator.
\foot{Spin projection matrices were originally introduced in
refs.~\refs{\Kuhn,\Guberina}.  The $\ubar$ and $v$ spinors in our definition
of $P_{S S_z}$ correspond to an outgoing $Q\Qbar$ pair.  They differ from the
$\vbar$ and $u$ spinors appearing in these earlier articles' spin projection
operators.  Our $(+,-,-,-)$ metric signature convention is also opposite to
that of ref.~\Guberina.}
This $4\times 4$ matrix reduces to the covariant expressions
\eqn\spinprojopII{\eqalign{
P_{00}(P;q) &= {-1 \over 2\sqrt{2} \MQ} \bigl( {\Pslash\over 2} - \qslash
- \MQ \bigr) \gfive \bigl({\Pslash\over 2}+\qslash+ \MQ \bigr) \cr
P_{1 S_z}(P;q) &= {-1 \over 2\sqrt{2} \MQ} \bigl({\Pslash\over 2} - \qslash
- \MQ \bigr) \slash{\varepsilon}^*(P;S_z)
\bigl({\Pslash\over 2}+\qslash+ \MQ \bigr) \cr}}
up to $O(q^2)$ corrections.  The projection operators' dependence on
$q$ is irrelevant for S-wave $Q\Qbar$ production.  The $g^* \to \QQbaroctet$
subamplitude consequently simplifies to
\eqn\goctetampIII{\CA(g_a^* \to \QQbaroctet_b)_\mu = \gS M
\varepsilon_\mu^*(P;S_z) \d_{ab}}
where $M=2 \MQ$ denotes the mass of the heavy quark-antiquark pair.  Combining
this result with the remaining perturbative part of the Feynman graph in
\qqbaroctet, we determine the short distance matrix element in
eqn.~\qqbarpsiQamp:
\eqn\qqbaroctetamp{ \CA(q(p_1) \qbar(p_2) \to \QQbaroctet_a)_{\rm short \>
distance} = {\gS^2 \over M} \vbar(p_2) \gamma^\mu T_a u(p_1)
\varepsilon^*_\mu(p_1+p_2;S_z).}

	Once the color-octet $Q\Qbar$ pair has been formed, it can evolve into
a physical $\psiQ$ bound state in many different ways.  For example, it may
emit two long wavelength gluons in a double chromoelectric dipole transition
and turn into a color-singlet $Q\Qbar$ pair that has order unity overlap with
the full $\psiQ$ wavefunction in \psiwavefunc.  Alternatively, it may absorb a
soft gluon and transform into a $\chiQJ$ through the $O(v)$ Fock state in
\chiwavefunc.  The P-wave quarkonium can subsequently emit a photon and
decay to the S-wave state.  All possible ways to produce a $\psiQ$
starting from a $\QQbaroctet$ pair are encoded into the long distance
amplitude in eqn.~\qqbarpsiQamp.  This factor is pictorially represented
by the big blob in \qqbaroctet\ with the long wavelength gluon tail.  It
cannot be computed from first principles without resorting to lattice
QCD.  But since the low energy dependence of interactions that take place
inside the blob is negligible, the nonperturbative amplitude is simply a
constant:
\eqn\longamp{\CA\bigl(\QQbaroctet \to \psiQ(nS)\bigr)_{\rm long \> distance}
= \> {\rm const}.}
The total color-octet contribution to $\psiQ$ production thus involves one
free parameter which we will eventually have to fit.

	We turn now to investigate the gluon fusion process $gg \to \psiQ$.
The color-octet amplitude for this reaction again factorizes into short and
long distance pieces:
\eqn\ggpsiQamp{\CA\bigl(gg \to \psiQ(nS)\bigr)_{\rm octet}
= \CA(gg \to \QQbaroctet)_{\rm short \> distance} \times
\CA\bigl(\QQbaroctet \to \psiQ(nS)\bigr)_{\rm long \> distance}.}
The long distance term is precisely the same as in eqn.~\longamp.  The short
distance factor involves at lowest order the three Feynman diagrams displayed
in \ggoctet.  Evaluating the high energy gluonic matrix element in a similar
fashion to its quark counterpart in \qqbaroctetamp, we deduce
\eqn\ggoctetamp{\eqalign{\CA(g_a(p_1) g_b(p_2) \to \QQbaroctet_c)_{\rm short \>
distance} &= {i\gS^2 \over M} f_{abc} \varepsilon^\mu(p_1) \varepsilon^\nu(p_2)
\varepsilon^\sigma(p_1+p_2;S_z)^* \cr
\times {p_1^2 + p_2^2 \over p_1^2 + p_2^2 -M^2}  &
\bigl[ (p_2-p_1)_\sigma g_{\u\v} + 2 (p_{1\v} g_{\u\o} - p_{2\u} g_{\v\o})
\bigr]. \cr}}
It is important to observe that this expression vanishes when both incoming
gluons are on-shell.  This result is consistent with Yang's theorem which
forbids a massive $J=1$ vector boson from decaying to two massless $J=1$
bosons \Yang.  In fact, the theorem requires that the on-shell
$gg \to \psiQ$ amplitude vanish to all orders.  The color-octet diagrams
in \ggoctet\ consequently participate in quarkonium production only as
subgraphs inside more complicated reactions.

	$O(\aS^2)$ scattering processes create quarkonia at small transverse
momenta that differ from zero only as a result of the incident partons'
intrinsic hadronic motion.  Since outgoing states from $2 \to 1$ collisions
are lost down the beampipe, we need to consider $2 \to 2$ transitions which
produce quarkonia with experimentally measurable $\pT$.  Such reactions start
at $O(\aS^3)$ and proceed in both the color-singlet and color-octet mechanisms
through the partonic channels $q\qbar \to Q\Qbar[^{2S+1}L_J^{(C)}]g$,
$gq \to Q\Qbar[^{2S+1}L_J^{(C)}]q$ and $gg \to Q\Qbar[^{2S+1}L_J^{(C)}]g$.
The differential cross sections for the color-singlet modes were calculated in
refs.~\refs{\Baier{--}\Humpert}.  We present their color-octet analogues
below.

	The Feynman diagrams which generate $\psiQ$ quarkonia in the three
different partonic modes through the color-octet mechanism are illustrated in
figs.~5a, 5b and 5c.  The shaded circles in these figures represent the
$q\qbar \to \psiQ$ and $gg \to \psiQ$ amplitudes in eqns.~\qqbarpsiQamp\ and
\ggpsiQamp.  The sums of the graphs in each channel yield gauge invariant
amplitudes.  Their spin and color averaged squares are given by
\eqna\sqrdamps
$$ \eqalignno{
& \mathop{{\bar{\sum}}} | \CA(q\qbar \to \psiQ g) |^2 =
{512 \pi^3 \aS^3 \over 27 M^2} {\that^2+\uhat^2 +2M^2 \shat \over
\that\uhat (\shat-M^2)^2} \bigl[4(\that^2+\uhat^2)-\that\uhat\bigr]
\longmatrixelement \qquad & \sqrdamps a \cr
& \mathop{{\bar{\sum}}} | \CA(gq \to \psiQ q) |^2 =
-{64 \pi^3 \aS^3  \over 9 M^2} {\shat^2+\uhat^2+2M^2\that \over
\shat\uhat(\that-M^2)^2} \bigl[4(\shat^2+\uhat^2)-\shat\uhat \bigr]
\longmatrixelement \qquad & \sqrdamps b \cr
& \mathop{{\bar{\sum}}} | \CA(g g \to \psiQ g) |^2 =
-{32 \pi^3 \aS^3  \over 3 M^2} {27(\shat\that+\that\uhat+\uhat\shat)
-19M^4 \over \bigl[(\shat-M^2)(\that-M^2)(\uhat-M^2)\bigr]^2} & \cr
& \quad \times
\bigl[(\that^2 +\that\uhat+\uhat^2)^2 - M^2 (\that +\uhat)
(2\that^2+\that\uhat+2\uhat^2)+M^4(\that^2+\that\uhat+\uhat^2) \bigr]
\longmatrixelement \qquad & \sqrdamps c \cr} $$
where $\longmatrixelement \equiv |\CA \bigl(\QQbaroctet \to
\psiQ(nS)\bigr)|^2$.  The squared matrix elements
in \sqrdamps{a}\ and \sqrdamps{b}\ were determined using standard spinor
summation techniques.  The result in \sqrdamps{c}\ was obtained from a simple
helicity amplitude method which obviated the need to compute a large number of
interference terms.  We calculated the 24 different helicity amplitudes
for $gg \to \psiQ g$ scattering by choosing explicit representations for
the massless and massive vector boson momenta and polarizations in the parton
center-of-mass \break frame.
\foot{Parity and crossing symmetry relations imply that only 4 of the 24
$gg \to \psiQ g$ helicity amplitudes are independent.}
The amplitudes were then individually squared and added together using the
high energy physics Mathematica package FEYNCALC \Mertig.

	The squared matrix element in eqn.~\sqrdamps{a}\ diverges when either
of the partonic Mandelstam invariants $\that$ or $\uhat$ approach zero.  The
source of these long distance infinities can be identified by looking at
\octetgraphs{a}.  As the outgoing gluon in a $q\qbar \to \psiQ g$
collision becomes collinear with either the incoming quark or antiquark, the
exchanged fermion in the $t$ or $u$-channel graphs goes on shell and its
propagator blows up.  The divergence generated by the collinear splitting of
the incident parton may be factored into its distribution function.  The hard
scattering reaction then reduces to the $q\qbar \to \psiQ$ Born level
process illustrated in \qqbaroctet.  A similar $u$-channel collinear
divergence develops in the squared $gq \to \psiQ q$ amplitude when the
four-momenta of the incoming gluon and outgoing quark become proportional.  In
contrast, the $t$-channel diagram in \octetgraphs{b}\ remains finite.  The
pole in the intermediate propagator is canceled by the zero in the
$gg \to \psiQ$ amplitude as the $t$-channel gluon goes on shell.  This same
cancellation renders $\mathop{{\bar{\sum}}} | \CA(g g \to \psiQ g) |^2$ free of
long distance infinities.

	The squared color-octet amplitudes in \sqrdamps{}\ enter into the
partonic differential cross section
\eqn\partonxsect{ {d\sigma\over d\that} (ab \to \psiQ c)_{\rm octet} =
{1 \over 16 \pi \shat^2} \mathop{{\bar{\sum}}}
| \CA \bigl(ab \to \psiQ c \bigr) |^2}
which appears in turn within the triple differential hadronic expression
\eqn\hadronxsect{ {d^3 \sigma \over dy_3 dy_4 d\pT}(AB \to \psiQ X)_{\rm
octet}= 2\pT \sum_{abc} x_a x_b f_{a/A}(x_a) f_{b/B}(x_b) {d\sigma\over d\that}
(ab \to \psiQ c)_{\rm octet}.}
The partonic cross section is multiplied by distribution functions
$f_{a/A}(x_a)$ and $f_{b/B}(x_b)$ that specify the probability of finding
partons $a$ and $b$ inside hadrons $A$ and $B$ carrying momentum fractions
$x_a$ and $x_b$.  The product is then summed over parton channels.
The resulting hadronic cross section is a function of the $\psiQ$ and
recoiling jet rapidities $y_3$ and $y_4$ and their common transverse momentum
$\pT$.

	It is instructive to examine the high energy limit of color-octet
quarkonia production.   As the partonic Mandelstam invariants grow to
infinity, the cross section in eqn.~\partonxsect\ reduces to
\eqn\highenergyxsect{{d\sigma \over d\that}(ab \to \psiQ c)_{\rm octet}
\sp {\buildrel \shat \to \infty \over \longrightarrow} \sp
{d\sigma\over d\that}(ab \to g^* c) \times \Bigl({1 \over M^2} \Bigr)^2 \times
|\CA \bigl(g^* \to \psiQ \bigr)|^2.}
This asymptotic expression has a simple gluon fragmentation interpretation.
The first factor represents the differential cross section for producing a
high energy virtual gluon.  The second term comes from the square of the
gluon's propagator.  The last factor equals the square of the amplitude in
eqn.~\goctetampIII\ times $\longmatrixelement$ and determines the virtual
gluon's probability to hadronize into a $\psiQ$ bound state.  The gluon
fragmentation picture for heavy quarkonium production is thus precisely
recovered in the high energy limit \BraatenYuanI.

	Gluon fragmentation via the color-octet mechanism represents the
dominant source of large $\pT$ quarkonia at hadron colliders
\refs{\BraatenYuanII,\BDFM,\Cho}.  The total cross section for $\psiQ$
production reduces at high energies to the fragmentation form
\eqn\fragxsect{{d^3\sigma \over dy_3 dy_4 d\pT}
\bigl(AB \to \psiQ X \bigr)_{\rm frag} = \int_0^1 dz \,
{d^3\sigma \over dy_3 dy_4 d\pT}\bigl(AB \to g\bigl({\pT \over z}\bigr) X,\mu
\bigr) \> D_{g \to \psiQ}(z,\mu).}
The gluon fragmentation function evaluated at the factorization scale $\mu=M$
is readily identified from eqn.~\highenergyxsect:
\eqn\glufragfunc{D_{g \to \psiQ}(z,M) = {4 \pi \aS(M) \longmatrixelement \over
M^2} \d(1-z).}
Leading log QCD corrections to this result may be summed up using the
Altarelli-Parisi equation
\eqn\APeqn{\mu {d D_{g \to \psiQ} \over d\mu}(z,\mu) = {\aS(\mu) \over
\pi} \int_z^1 {dy \over y} P_{gg}(y) D_{g \to \psiQ} \bigl({z \over y},\mu
\bigr)}
where
\eqn\split{P_{gg}(y) = 6 \Bigl[ {y \over (1-y)_+} + {1-y \over y} + y(1-y) +
{33 -2\nf \over 36} \d(1-y) \Bigr]}
denotes the gluon splitting function for $\nf$ active quark flavors.  At high
energies, the fragmentation approximation in \fragxsect\ incorporates
sizable $O(\log(E^2/M^2))$ renormalization effects, and its intrinsic
$O(M^2/E^2)$ errors are negligible.  In contrast, the color-octet formula
in \hadronxsect\ does not include any QCD corrections which are small
at low $\pT$, but it retains full dependence upon all $O(M^2/E^2)$ terms.
These two forms for the $\psiQ$ differential cross section are thus
complementary.

	We may exercise our perturbative freedom to supplement the $O(\aS^3)$
color-octet cross section with the leading logarithms from the gluon
fragmentation approach.  We implement this choice as follows:
\eqn\interpxsect{d \sigma \bigl(AB \to \psiQ X \bigr)_{\rm interp}
= d\sigma \bigl(AB \to \psiQ X \bigr)_{\rm octet}
 \times { d\sigma \bigl(AB \to \psiQ X \bigr)_{\rm frag \sp with \sp QCD \sp
          running}
 \over    d\sigma \bigl(AB \to \psiQ X \bigr)_{\rm frag \sp without \sp
          QCD \sp running} }. }
The ratio on the RHS of this hybrid expression reduces to unity at low
energy, whereas the first factor over the denominator approaches unity at
high energy.  Eqn.~\interpxsect\ therefore smoothly interpolates between the
two asymptotic limits in eqns.~\hadronxsect\ and \fragxsect.   We will use
this final color-octet formula to study bottomonia and charmonia production at
the Tevatron in the following section.

\newsec{Upsilon and Psi production at the Tevatron}

	The CDF collaboration has recently measured Upsilon production at the
Tevatron for the first time \Upsilondata.   Their reported $\Upsilon(1S)$,
$\Upsilon(2S)$ and $\Upsilon(3S)$ transverse momentum differential
cross sections seriously disagree with predictions based upon $O(\aS^3)$
color-singlet cross section formulae.  While some discrepancy between the
theoretical and experimental findings may be attributed to various parameter
uncertainties and higher order corrections, the magnitude of the disparity
strongly suggests that additional production mechanisms beyond the simplest
color-singlet processes are at work.  As we shall see, inclusion of
color-octet reactions brings the theoretical cross sections into line with
the Upsilon data.

	We first investigate the differential cross section for $\Upsilon(1S)$
production in the rapidity range $|y| \le 0.4$.  The color-singlet
distribution is illustrated by the dashed curve in \break
\UpsilononeSxsect.
\foot{The differential cross sections displayed in \UpsilononeSxsect\ and
subsequent figures were calculated using the MRSD0 parton distribution
functions.}
It includes direct $\Upsilon(1S)$ production as well as radiative feeddown
{}from $\chibJ(1P)$ and $\chibJ(2P)$ states.  Additional contributions from
strong and electromagnetic decays of $\Upsilon(2S)$ and $\Upsilon(3S)$ are
negligible.  In computing the color-singlet cross sections, we adopted
the radial wavefunction values tabulated in ref.~\Quigg\ for the
Buchm\"uller-Tye QCD potential \Buchmuller\ and the associated bottom quark
mass $\Mb = 4.88 \GeV$.   The color-singlet prediction for $\Upsilon(1S)$
production clearly underestimates the CDF data points shown in the figure.

	The dot-dashed curve in \UpsilononeSxsect\ displays the color-octet
contribution to the $\Upsilon(1S)$ differential cross section.  The
best fit value $\CM_8\bigl(\Upsilon(1S)\bigr) = (3.66 \pm 0.27) \times 10^{-3}
\GeV^2$ for its squared long distance matrix element determines the probability
for a $b\bbar[\octet]$ pair to hadronize into an $\Upsilon(1S)$.  Following
eqn.~\interpxsect, we have incorporated leading log QCD corrections into the
color-octet cross section.  The dot-dashed curve therefore approaches by
construction the dotted gluon fragmentation curve based upon the same value for
$\CM_8\bigl(\Upsilon(1S)\bigr)$ at high $\pT$.  The two cross sections diverge,
however, at transverse momenta comparable to the $\Upsilon(1S)$ mass.  This
disparity demonstrates the breakdown of the fragmentation picture in the low
energy regime.

	The sum of the color-singlet and color-octet transverse momentum
distributions is illustrated by the solid curve in \UpsilononeSxsect.
It represents the total theoretical $\Upsilon(1S)$ differential cross
section.  The solid curve's shape fits the data quite well except at the
lowest point.  At very small $\pT$, the color-singlet and color-octet
cross sections are corrupted by collinear divergences as discussed in
section 2.  Since the intrinsic motion of incident partons inside colliding
hadrons renders the differential cross section uncertain for $\pT \ltap 2
\GeV$, we have not attempted to remove the divergences from the dashed and
dot-dashed curves.  Instead, we simply regard the portion of the plot which
runs below $\pT = 2 \GeV$ as untrustworthy.  Above this low momentum
cutoff, the predicted $\Upsilon(1S)$ distribution matches the data.

	Analogous theoretical and experimental differential cross sections for
$\Upsilon(2S)$ production in the rapidity region $|y| \le 0.4$ are displayed in
\UpsilontwoSxsect.  The dashed color-singlet curve in the figure
incorporates direct $\Upsilon(2S)$ production and radiative feeddown from
$\chibJ(2P)$.  The dot-dashed color-octet and dotted gluon fragmentation
curves correspond to the squared matrix element $\CM_8\bigl(\Upsilon(2S)\bigr)
= (2.25 \pm 0.38) \times 10^{-3} \GeV^2$.  The solid curve again illustrates
the combined color-singlet and color-octet contributions to the $\Upsilon(2S)$
differential cross section.  It clearly agrees with the CDF measurements much
better than the color-singlet distribution alone.

	We consider next the $\Upsilon(3S)$ differential cross section for
which data is limited but interesting.  The extent to which radiative
$\chibJ(3P) \to \Upsilon(3S) \gamma$ transitions mediate $\Upsilon(3S)$
formation is unknown, for the $n=3$ P-wave states have never been observed.
Although the $\Upsilon(4S)$ can electromagnetically decay to them, it lies
above the $B\Bbar$ threshold.  As a result, $\Br(\Upsilon(4S) \to \chibJ(3P)
\gamma)$ is very small,
\foot{CLEO has recently set a 5\% upper limit on non-$B\Bbar$ decays of the
$\Upsilon(4S)$ \CLEO.}
and the $n=3$ P-wave wave levels are rarely populated at $e^+ e^-$ colliders.
But at the Tevatron, $\chibJ(3P)$ bottomonia are directly created in
hadronic reactions.  Since the integrated $\Upsilon(3S)$ cross section measured
by CDF is 40 times larger than the color-singlet model's prediction when only
$\Upsilon(3S)$ production is taken into account, it is believed that $n=3$
P-wave levels reside below the $B\Bbar$ threshold and decay at a significant
rate to the $n=3$ S-wave state \Upsilondata.  We consequently include their
contributions into the dashed color-singlet curve in \UpsilonthreeSxsect\
using the measured $n=2$ branching ratios $\Br(\chibJ(2P) \to \Upsilon(2S)
\gamma)$ as estimates for their unknown $n=3$ counterparts.  The dot-dashed,
dotted and solid curves in the figure represent the color-octet, gluon
fragmentation and total differential cross sections based upon
$\CM_8\bigl(\Upsilon(3S)\bigr) = (0.53 \pm 0.24) \times 10^{-3} \GeV^2$.
Given all the uncertainties, we can only say that agreement between the few
measured $\Upsilon(3S)$ data points and the theoretical prediction appears
adequate.

	We now turn to the charmonium sector and examine prompt $\Jpsi$
production at the Tevatron within the pseudorapidity interval $|\eta| \le
0.6$.  The $\Jpsi$ cross section multiplied by the muon branching fraction
$\Br(\Jpsi \to \mu^+ \mu^-)=0.0597$ is illustrated in \Jpsixsect.  The dashed
curve in the figure depicts the color-singlet differential cross section
that includes direct $\Jpsi$ production and radiative feeddown from
$\chicJ(1P)$ states.  In calculating this curve, we employed the
Buchm\"uller-Tye charmonium wave function values at the origin tabulated in
ref.~\Quigg\ and the associated charm quark mass $\Mc=1.48 \GeV$.  Additional
color-singlet contributions to the $\Jpsi$ cross section from $\psi'$
production are small.  The dot-dashed curve in \Jpsixsect\ shows the
color-octet $\Jpsi$ cross section for the squared matrix element value
$\CM_8\bigl(\Jpsi\bigr) = (0.68 \pm 0.03) \times 10^{-3} \GeV^2$.  It is
almost identical to the solid curve which represents the sum of the
color-singlet and color-octet differential distributions.  Comparing the
dot-dashed color-octet and dotted gluon fragmentation curves with the CDF data
in \Jpsixsect, we see that the fragmentation approximation appears to fit the
experimental measurements better at small transverse momenta.  But we must
stress that this low energy agreement is fortuitous.  The fragmentation
prediction looks better only in the region where it becomes unreliable.

	We finally consider prompt $\psi'$ production at the Tevatron.
The observed differential cross section is 30 times larger than recent
theoretical predictions based upon $c \to \psi'$ and $g \to \psi'$
fragmentation calculations \refs{\BDFM,\Roy}.  These two $O(v^3)$
fragmentation processes create color-singlet $c\cbar[^3S_1^{(1)}]$ pairs at
short distance scales and yield comparable $\psi'$ production rates.  At
higher order in the NRQCD velocity expansion but lower order in perturbative
QCD, the color-octet mechanism also mediates $g \to \psi'$ fragmentation
through an intermediate $c\cbar[\octet]$ pair.  The additional color-octet
contribution to the $\psi'$ cross section which was not previously taken into
account can readily resolve the factor of 30 discrepancy between theory and
data \refs{\ChoWise,\BraatenFleming}.

	We plot in \Psipxsect\ the $\psi'$ differential cross section
multiplied by the muon branching fraction $\Br(\psi' \to \mu^+ \mu^-)=0.0077$.
A pseudorapidity cut $|\eta| \le 0.6$ has been applied to the distributions
shown in the figure.  The dashed curve represents just direct $\psi'$
production, for no other charmonium level below the $D\Dbar$ threshold can
decay to the $n=2$ state.  This color-singlet prediction drastically
underestimates the measured differential cross section.   Inclusion of
color-octet contributions with $\CM_8\bigl(\psi'\bigr) = (0.20 \pm 0.01)
\times 10^{-3} \GeV^2$ brings the magnitude and shape of the theoretical
distribution into agreement with the data points displayed in \Psipxsect.
The color-octet mechanism thus appears to solve the CDF $\psi'$ problem.

\newsec{NRQCD matrix elements}

	We have seen that heavy quarkonia production involves short and long
distance physics.  The high energy creation of color-singlet and color-octet
$Q\Qbar$ pairs is perturbatively computable within the NRQCD effective theory.
The subsequent low energy hadronization of these pairs into physical bound
states is described in terms of nonrenormalizable operator matrix elements.
NRQCD scaling rules determine the relative importance of different long
distance matrix elements and yield relations among them.  We will use these
rules to check the consistency of the fitted Upsilon and Psi amplitude values
which we obtained in the preceding section.

	We first recall the relations
\eqn\singletmatrixelements{\eqalign{
\langle 0 | \CO_1^{\psiQ}(^3S_1) | 0 \rangle &= {N_c \over 2 \pi} |R(0)|^2
= O(\MQ^3 \vQ^3) \cr
\langle 0 | \CO_1^{\chiQJ}(^3P_J) | 0 \rangle &= {3 N_c \over 2 \pi} (2J+1)
 |R'(0)|^2 = O(\MQ^5 \vQ^5) \cr}}
between quarkonia radial wavefunctions at the origin and matrix elements of
certain color-singlet four-quark operators in the NRQCD Lagrangian \Bodwin.
The details of these operators' definitions are discussed in ref.~\Bodwin, but
they are not important to us.  Instead, we are only interested in their
scaling dependence upon the heavy quark mass $\MQ$ and velocity $\vQ$.

	Using the Buchm\"uller-Tye wavefunction information tabulated in
ref.~\Quigg, we list the numerical values of color-singlet matrix elements
that are relevant for Upsilon and Psi production in Table I.
\eject
$$ \vbox{\offinterlineskip
\def\tablerule{\noalign{\hrule}}
\hrule
\halign {\vrule#& \strut#&
\ \hfil#\hfil & \vrule#&
\ \hfil#\hfil & \vrule#&
\ \hfil#\hfil & \vrule# \cr
\tablerule%
height10pt && \omit && \omit && \omit &\cr
&& Color-Singlet && Numerical && NRQCD &\cr
&& Matrix Element && Value &&\quad Scaling Order\quad&\cr
height10pt && \omit && \omit && \omit & \cr
\tablerule
height10pt && \omit && \omit && \omit &\cr
&& \quad $\langle 0|O_1^{J/\psi}({}^3S_1)|0\rangle$ \quad &&
\quad $3.9\times10^{-1}\ {\rm GeV}^3$\quad && \quad $\Mc^3\vc^3$ \quad &\cr
height10pt && \omit && \omit && \omit &\cr
&& \quad $\langle 0|O_1^{\chi_{c1}}({}^3P_1)|0\rangle$ \quad &&
\quad $3.2\times10^{-1}\ {\rm GeV}^5$\quad && \quad $\Mc^5\vc^5$ \quad &\cr
height10pt && \omit && \omit && \omit &\cr
&& \quad $\langle 0|O_1^{\psi'}({}^3S_1)|0\rangle$ \quad &&
\quad $2.5\times10^{-1}\ {\rm GeV}^3$ \quad && \quad $\Mc^3\vc^3$ \quad &\cr
height10pt && \omit && \omit && \omit &\cr
&& \quad $\langle 0|O_1^{\Upsilon(1S)}({}^3S_1)|0\rangle$ \quad &&
\quad $3.1\ {\rm GeV}^3$ \quad && \quad $\Mb^3\vb^3$ \quad &\cr
height10pt && \omit && \omit && \omit &\cr
&& \quad $\langle 0|O_1^{\chi_{b1}(1P)}({}^3P_1)|0\rangle$ \quad &&
\quad $6.1\ {\rm GeV}^5$\quad && \quad $\Mb^5\vb^5$ \quad &\cr
height10pt && \omit && \omit && \omit &\cr
&& \quad $\langle 0|O_1^{\Upsilon(2S)}({}^3S_1)|0\rangle$ \quad &&
\quad $1.5\ {\rm GeV}^3$ \quad && \quad $\Mb^3\vb^3$ \quad &\cr
height10pt && \omit && \omit && \omit &\cr
&& \quad $\langle 0|O_1^{\chi_{b1}(2P)}({}^3P_1)|0\rangle$ \quad &&
\quad $7.1\ {\rm GeV}^5$\quad && \quad $\Mb^5\vb^5$ \quad &\cr
height10pt && \omit && \omit && \omit &\cr
&& \quad $\langle 0|O_1^{\Upsilon(3S)}({}^3S_1)|0\rangle$ \quad &&
\quad $1.2\ {\rm GeV}^3$ \quad && \quad $\Mb^3\vb^3$ \quad &\cr
height10pt && \omit && \omit && \omit &\cr
&& \quad $\langle 0|O_1^{\chi_{b1}(3P)}({}^3P_1)|0\rangle$ \quad &&
\quad $7.7\ {\rm GeV}^5$\quad && \quad $\Mb^5\vb^5$ \quad &\cr
height10pt && \omit && \omit && \omit &\cr
\tablerule}} $$
\centerline{Table I}
\medskip\noindent
The entries in this table are consistent with the NRQCD power counting rules.
In the $\vQ \to 0$ limit, the long distance quarkonia matrix elements would be
independent of the radial quantum number $n$.  In actuality, the $n$
dependence of the $b\bbar$ sector entries is generally smaller than that of
the $c\cbar$ values.  This trend simply indicates that the NRQCD velocity
expansion works better for bottomonia with $\vb^2 \simeq 0.08$ than for
charmonia with $\vc^2 \simeq 0.23$.

	NRQCD scaling rules can also be applied to color-octet four quark
operators.  We focus upon the matrix elements
\eqn\octetmatrixelements{\eqalign{
\langle 0 | \CO_8^{\psiQ}(^3S_1) | 0 \rangle &= O(\MQ^3 \vQ^7) \cr
\langle 0 | \CO_8^{\chiQJ}(^3S_1) | 0 \rangle &= O(\MQ^3 \vQ^5) \cr}}
that determine the probability for $\QQbaroctet$ pairs to hadronize into
$\psiQ$ and $\chiQJ$ quarkonia.  These matrix elements appear in the
decomposition of the total long distance squared amplitude
\eqn\octetdecomp{\eqalign{\CM_8 \bigl(\psi_Q(n)\bigr) = {1 \over 24 \MQ}
\sum_{m \ge n} \Bigl\{ & \langle 0 | \CO_8^{\psiQ(m)}(^3S_1) | 0 \rangle \;
\Br \bigl(\psiQ(m) \to \psiQ(n) \bigr)\cr
& + \sum_{J=0,1,2} \langle 0 | \CO_8^{\chiQJ(m)}(^3S_1) | 0 \rangle \;
 \Br\bigl(\chiQJ(m) \to \psiQ(n) \gamma \bigr) + \cdots \Bigr\} \cr}}
which reflects the different ways for a color-octet pair to materialize as a
final $\psiQ$.  The infinite number of terms represented by the ellipsis
contribute at higher order in the $\vQ$ or $\aS$ expansions to color-octet
$\psiQ$ production than those explicitly displayed.  So we will neglect them
for simplicity.

	We can extract numerical values for the NRQCD color-octet matrix
elements in eqn.~\octetmatrixelements\ by inserting our fitted values for
$\CM_8(\psiQ)$ into eqn.~\octetdecomp.  Combining $\CM_8(\Jpsi) =
6.8 \times 10^{-4} \GeV^2$ and $\CM_8(\psi')=2.0 \times 10^{-4} \GeV^2$ with
the recent CDF finding that 32.3\% prompt $\Jpsi$'s originate from $\chicJ$
decays \Bauer, we deduce the charmonium color-octet matrix elements
\eqna\matrixelements
$$ \eqalignno{
\langle 0 | \CO_8^{\Jpsi}(^3S_1) | 0 \rangle &= 1.2 \times 10^{-2} \GeV^3 &
\matrixelements a \cr
\langle 0 | \CO_8^{\psi'}(^3S_1) | 0 \rangle &= 7.3 \times 10^{-3} \GeV^3 &
\matrixelements b \cr
\langle 0 | \CO_8^{\chi_{c1}}(^3S_1) | 0 \rangle &= 1.6 \times 10^{-2} \GeV^3.
& \matrixelements c \cr} $$
In the absence of additional bottomonium experimental information, we employ
the NRQCD relation
\eqn\NRQCDreln{{\langle 0 | \CO_8^{\Upsilon(nS)}(^3S_1) | 0 \rangle \over
\langle 0 | \CO_8^{\psi(nS)}(^3S_1) | 0 \rangle} \simeq \Bigl( {\Mb\over\Mc}
\Bigr)^3 \Bigl({\vb\over\vc}\Bigr)^7}
to scale the Psi values in \matrixelements{a}\ and \matrixelements{b}\
to the Upsilon sector.  We then obtain the color-octet matrix elements listed
in Table II.
\eject
$$ \vbox{\offinterlineskip
\def\tablerule{\noalign{\hrule}}
\hrule
\halign {\vrule#& \strut#&
\ \hfil#\hfil & \vrule#&
\ \hfil#\hfil & \vrule#&
\ \hfil#\hfil & \vrule# \cr
\tablerule%
height10pt && \omit && \omit && \omit &\cr
&& Color-Octet && Numerical && NRQCD &\cr
&& Matrix Element && Value &&\quad Scaling Order\quad&\cr
height10pt && \omit && \omit && \omit & \cr
\tablerule
height10pt && \omit && \omit && \omit &\cr
&& \quad $\langle 0|O_8^{J/\psi}({}^3S_1)|0\rangle$ \quad &&
\quad $1.2\times10^{-2}\ {\rm GeV}^3$\quad && \quad $\Mc^3\vc^7$ \quad &\cr
height10pt && \omit && \omit && \omit &\cr
&& \quad $\langle 0|O_8^{\chi_{c1}}({}^3S_1)|0\rangle$ \quad &&
\quad $1.6\times10^{-2}\ {\rm GeV}^3$\quad && \quad $\Mc^3\vc^5$ \quad &\cr
height10pt && \omit && \omit && \omit &\cr
&& \quad $\langle 0|O_8^{\psi'}({}^3S_1)|0\rangle$ \quad &&
\quad $7.3\times10^{-3}\ {\rm GeV}^3$ \quad && \quad $\Mc^3\vc^7$ \quad &\cr
height10pt && \omit && \omit && \omit &\cr
&& \quad $\langle 0|O_8^{\Upsilon(1S)}({}^3S_1)|0\rangle$ \quad &&
\quad $9.5\times10^{-3}\ {\rm GeV}^3$ \quad && \quad $\Mb^3\vb^7$ \quad &\cr
height10pt && \omit && \omit && \omit &\cr
&& \quad $\langle 0|O_8^{\chi_{b1}(1P)}({}^3S_1)|0\rangle$ \quad &&
\quad $4.3\times10^{-1}\ {\rm GeV}^3$\quad && \quad $\Mb^3\vb^5$ \quad &\cr
height10pt && \omit && \omit && \omit &\cr
&& \quad $\langle 0|O_8^{\Upsilon(2S)}({}^3S_1)|0\rangle$ \quad &&
\quad $5.7\times10^{-3}\ {\rm GeV}^3$ \quad && \quad $\Mb^3\vb^7$ \quad &\cr
height10pt && \omit && \omit && \omit &\cr
&& \quad $\langle 0|O_8^{\chi_{b1}(2P)}({}^3S_1)|0\rangle$ \quad &&
\quad $5.2\times10^{-1}\ {\rm GeV}^3$\quad && \quad $\Mb^3\vb^5$ \quad &\cr
height10pt && \omit && \omit && \omit &\cr
\tablerule}} $$
\centerline{Table II}
\medskip

	The numerical entries in this second table should be regarded as
approximate estimates.  Nonetheless, we see that the NRQCD scaling rules hold
reasonably well for the color-octet matrix elements.  The magnitudes of the
entries in Table~II appear consistent with each other and with those in
Table~I.  These results thus provide a useful check on the entire color-octet
quarkonia production picture.

\newsec{Quarkonia spin alignment}

	We have so far not considered final state spin information in our
analysis.  Previous investigations have found that quarkonia generated at
large transverse momenta via gluon fragmentation processes like
$gg \to gg^* \to gg \chiQJ$ are significantly polarized \Cho.  The source of
this spin alignment can be traced to the fragmenting gluon.  At high energies,
the virtual $g^*$ is off-shell only by $O(M^2/E^2)$ and is therefore nearly
real and transverse.  The $\QQbaroctet$ pair which the gluon forms at short
distance scale must inherit its transverse alignment in order to conserve
angular momentum.  The subsequent long distance evolution of the color-octet
pair into a physical quarkonium state preserves the spins of the heavy quark
and antiquark up to $O(\Lambda_\QCD/M)$ corrections.  The final quarkonium's
polarization state is then fixed by heavy quark spin symmetry \ChoWise.

	It is interesting to examine the spin alignments of quarkonia produced
via the color-octet mechanism at low energies as well.  To do so, we decompose
the partonic squared amplitudes in eqn.~\sqrdamps{}\ into separate longitudinal
and transverse components:
\eqna\polsqrdamps
$$ \eqalignno{
\mathop{{\bar{\sum_{h=0}}}} | \CA(q\qbar \to \QQbarocteth g) |^2 &=
{512 \pi^3 \aS^3 \over 27 M^2} {4 M^2 \shat \over (\shat-M^2)^4}
\bigl[4(\that^2+\uhat^2)-\that\uhat \bigr]
& \polsqrdamps a \cr
\mathop{{\bar{\sum_{|h|=1}}}} | \CA(q\qbar \to \QQbarocteth g) |^2 &=
{512 \pi^3 \aS^3 \over 27 M^2} {\shat^2+M^4 \over (\shat-M^2)^4}
{\that^2+\uhat^2 \over \that\uhat} \bigl[4(\that^2+\uhat^2)-\that\uhat \bigr]
& \polsqrdamps b \cr
\mathop{{\bar{\sum_{h=0}}}} | \CA(gq \to \QQbarocteth q) |^2 &=
-{64 \pi^3 \aS^3 \over 9 M^2} {2 M^2 \that \over
\bigl[(\shat-M^2)(\that-M^2)\bigr]^2}
\bigl[4(\shat^2+\uhat^2)-\shat\uhat\bigr]
& \polsqrdamps c \cr
\mathop{{\bar{\sum_{|h|=1}}}} | \CA(gq \to \QQbarocteth q) |^2 &=
-{64 \pi^3 \aS^3 \over 9M^2} {(\shat^2+\uhat^2+2 M^2 \that)(\shat-M^2)^2
- 2 M^2 \shat\that\uhat \over \shat\uhat
\bigl[(\shat-M^2)(\that-M^2)\bigr]^2} & \cr
& \quad \times \bigl[4(\shat^2+\uhat^2)-\shat\uhat\bigr]
& \polsqrdamps d \cr
\mathop{{\bar{\sum_{h=0}}}} | \CA(g g \to \QQbarocteth g) |^2 &=
-{16 \pi^3 \aS^3 \over 3 M^2}
{2 M^2 \shat \over (\shat-M^2)^2} (\that^2+\uhat^2) \that\uhat & \cr
& \quad \times {27(\shat\that+\that\uhat+\uhat\shat)
-19M^4 \over \bigl[(\shat-M^2)(\that-M^2)(\uhat-M^2)\bigr]^2}
& \polsqrdamps e \cr
\mathop{{\bar{\sum_{|h|=1}}}} | \CA(g g \to \QQbarocteth g) |^2 &=
-{16 \pi^3 \aS^3 \over 3 M^2} {\shat^2 \over (\shat-M^2)^2}
\bigl[(\shat-M^2)^4+\that^4+\uhat^4+2 M^4 \bigl({\that\uhat \over \shat}
\bigr)^2 \bigr] & \cr
& \quad \times {27(\shat\that+\that\uhat+\uhat\shat)
-19M^4 \over \bigl[(\shat-M^2)(\that-M^2)(\uhat-M^2)\bigr]^2}
& \polsqrdamps f \cr} $$
The results in the $q\qbar \to \QQbarocteth g$ and $gq \to \QQbarocteth q$
channels were calculated using the covariant longitudinal and transverse spin
sums discussed in ref.~\Cho.  The $gg \to \QQbaroctet g$ squared matrix
elements were simply derived from our previously described helicity amplitude
method.  Summing over helicities in each mode, we reproduce the unpolarized
expressions in eqn.~\sqrdamps{}.

	The longitudinal squared amplitudes in \polsqrdamps{a},
\polsqrdamps{c}\ and \polsqrdamps{e}\ vanish as $\shat/M^2 \to \infty$,
whereas their transverse counterparts in \polsqrdamps{b}, \polsqrdamps{d}\ and
\polsqrdamps{f}\ approach constant limits.  The gluon fragmentation prediction
of total transverse alignment is therefore recovered at high energies.   This
asymptotic behavior can be seen in \Transratio\ where we plot the ratio
\eqn\polratio{
R_{\rm trans}= {\displaystyle{\sum_{|h|=1} {d\sigma \over d\pT}}
   (p\pbar \to \QQbarocteth X) \over
   \displaystyle{\sum_{h} {d\sigma \over d\pT}} (p\pbar \to \QQbarocteth X)}}
of transverse and total $\bbbaroctet$ and $\ccbaroctet$ differential cross
sections as a function of \break $\pT$.
\foot{The shapes of the $\bbbaroctet$ and $\ccbaroctet$ curves in
\Transratio\ are not exactly the same since their respective
$|y| \le 0.4$ rapidity and $|\eta| \le 0.6$ pseudorapidity cuts are
different.}
{}Fig.~11 also demonstrates that color-octet pairs are strongly transverse
even at low energies.  This surprising phenomenon has a simple explanation.
In the dominant $gg \to \QQbarocteth g$ channel, longitudinal color-octet pair
production along the beam axis is prohibited by angular momentum conservation.
The squared $h=0$ amplitude in \polsqrdamps{e}\ therefore vanishes as
$\that=-\half(\shat-M^2)(1-\cos\th^*)$ or
$\uhat=-\half(\shat-M^2)(1+\cos\th^*)$ approaches zero, while its $|h|=1$
counterpart in \polsqrdamps{f}\ remains finite.  Since small angle
scattering dominates this $2 \to 2$ partonic process, most $\QQbaroctet$ pairs
are formed at $|\cos\th^*| \simeq 1$.  As a result, they are transversely
aligned independent of their energy.  The ratios of the $h=0$ and $|h|=1$
squared amplitudes in the $q\qbar \to \QQbarocteth g$ and
$gq \to \QQbarocteth q$ channels similarly vanish as $\that\to 0$ or
$\uhat\to 0$.

	The nearly 100\% transverse alignment of $\ccbaroctet$ pairs should
be directly observable in $\psi'$ production at the Tevatron if the
color-octet mechanism resolves the $\psi'$ problem.  Color-octet
production then dominates over all other prompt $\psi'$ sources by more than a
factor of 30.  Heavy quark spin symmetry guarantees that $\psi'$ spin
alignments equal those of their $\ccbaroctet$ progenitors in the $\Mc
\to \infty$ limit \ChoWise.  Finite charm mass effects somewhat degrade
final $\psi'$ polarizations.  But experimental observation of a large
transverse $\psi'$ spin alignment would provide strong support for the
color-octet production picture.

	Other S and P-wave charmonia and bottomonia can also inherit a
sizable spin alignment.  Their helicity level populations depend upon the
color-octet matrix elements in Table II.  Detailed predictions for quarkonium
polarizations resulting from color-singlet and color-octet production will be
presented elsewhere \forthcoming.  But we emphasize here that experimental
measurements of Upsilon, Psi and Chi spin alignments will provide valuable
tests of our understanding of basic quarkonium physics.

\newsec{Conclusion}

	In this paper, we have examined color-octet production of heavy
quarkonia at hadron colliders.  Color-octet diagrams arise at the same order
in perturbative QCD as their color-singlet analogues and reduce to the dominant
set of gluon fragmentation graphs in the high energy limit.  They must
therefore be included in bottomonia and charmonia production computations.  As
we have seen, color-octet contributions to Upsilon and Psi differential cross
sections eliminate large disparities between earlier predictions and recent
measurements.  The long distance matrix element values needed to bring theory
into line with experiment are consistent with NRQCD scaling rules.
We have also found that the color-octet mechanism yields $\QQbaroctet$ pairs
which are nearly 100\% transversely aligned at all $\pT$.  Observation of
large quarkonia spin alignments would provide strong support for the
color-octet production picture.

	A number of extensions of this work would be interesting to pursue.
For example, a low transverse momentum study should reveal that quarkonia
distributions vanish rather than diverge as $\pT \to 0$.  In order to compute
the cross section turnover, it will be necessary to factorize collinear
singularities in both the color-singlet and color-octet channels into incident
parton distribution functions.  This factorization also needs to be performed
before differential quarkonia distributions can be integrated.  We expect
color-octet contributions to significantly enhance existing predictions for
total Upsilon and Psi production rates.  Comparison of integrated quarkonium
cross sections with fixed target measurements will provide an important means
for testing the long distance matrix element values which we have found using
collider data.  Finally, all the results in this article can be applied to the
study of quarkonia production at the next generation of hadron colliders.  We
look forward to further surprises in quarkonium physics coming from machines
like the LHC well into the next century.

\bigskip
\centerline{{\bf Acknowledgments}}
\bigskip

	We have benefitted from helpful discussions with many colleagues and
friends.  We wish to especially thank Eric Braaten, Michelangelo Mangano,
Miko\l aj Misiak, Ivan Muzinich, Vaia Papadimitriou, David Politzer, Chris
Quigg, Elizabeth Simmons, Sandip Trivedi and Mark Wise for sharing their
insights with us.

\listrefs
\listfigs

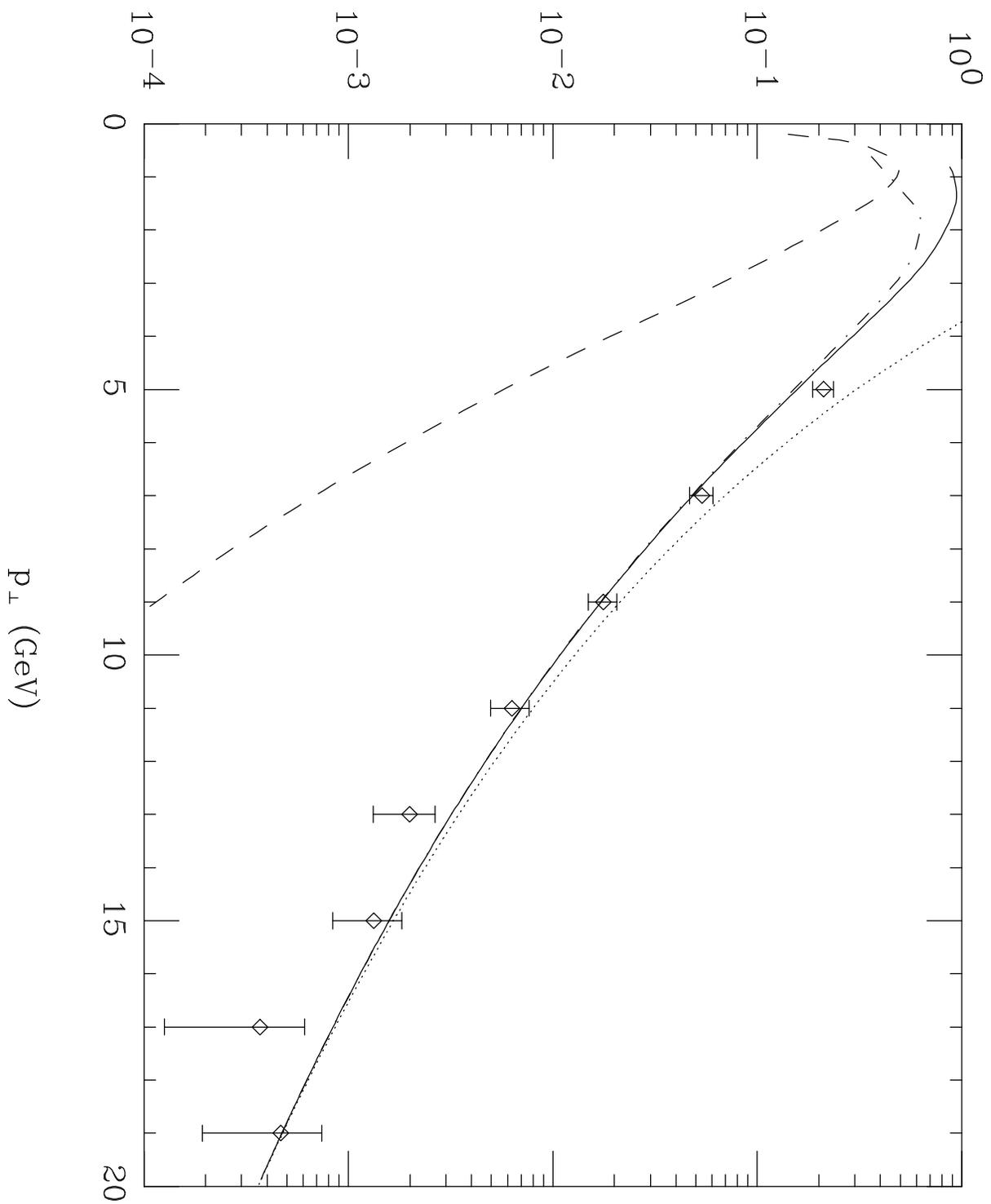
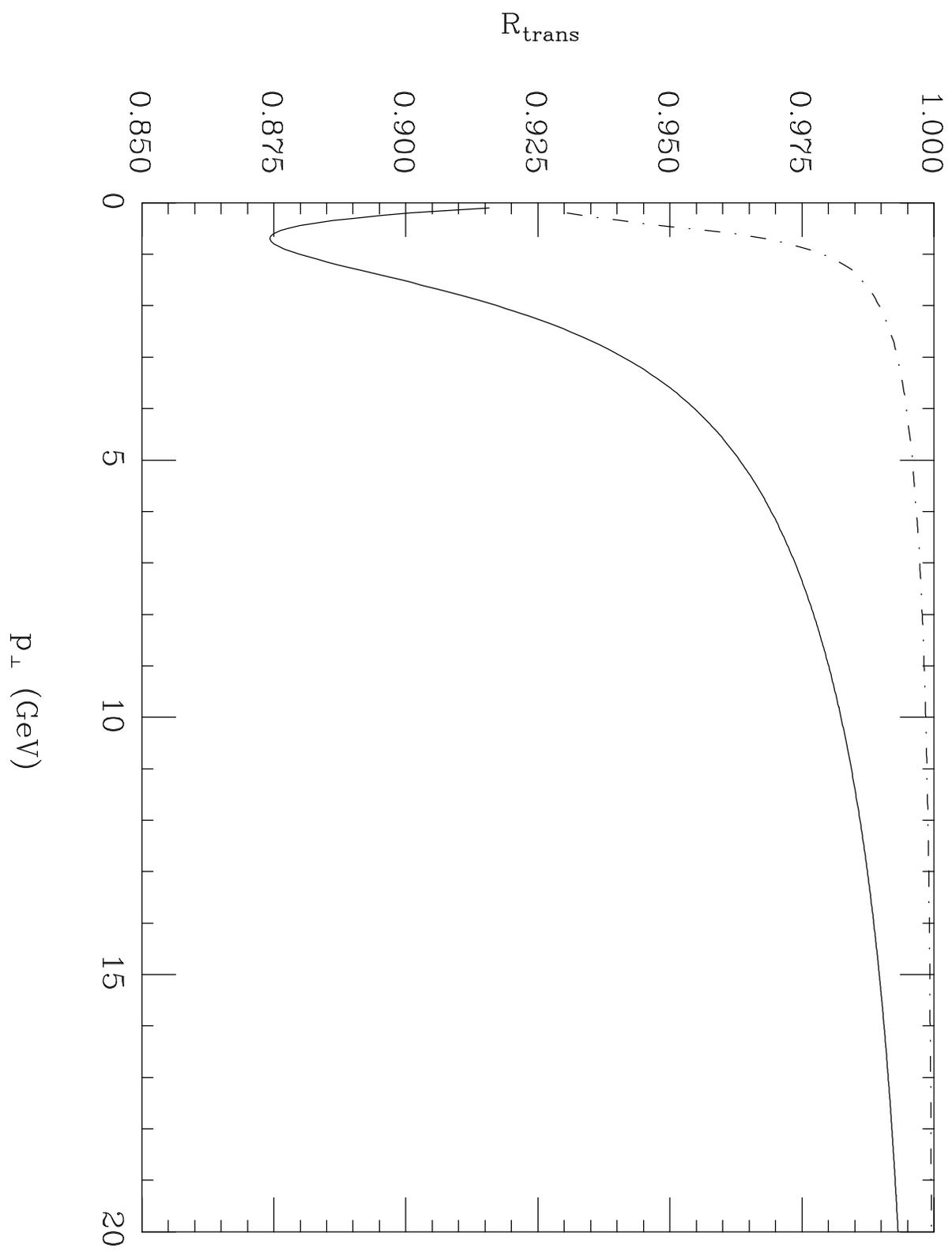
\bye